\documentstyle[aps,eqsecnum,epsfig]{revtex}

\newcommand{\bm}{\bibitem}
\newcommand{\tauiso}{{\mbox{\boldmath $\tau$}}}
\newcommand{\gmu}{{\gamma_\mu}}
\newcommand{\gf}{{\gamma_5}}

\tighten
\begin{document}

\draft

\twocolumn[\columnwidth\textwidth\csname@twocolumnfalse\endcsname

\title{ $pp \rightarrow pK^{+}\Lambda$ reaction in an
effective Lagrangian model }
\author{R. Shyam}
\address{ 
Saha Institute of Nuclear Physics, Calcutta 700064, India\\ 
E-mail address: shyam@tnp.saha.ernet.in}

\maketitle

\begin{abstract}
We investigate the $pp \rightarrow pK^{+}\Lambda$ reaction within
an effective Lagrangian model where the contributions to the 
amplitudes are taken into account within the tree level. The 
initial interaction between the two nucleons is modeled by 
the exchange of $\pi$, $\rho$, $\omega$ and $\sigma$ mesons and
the $\Lambda K^{+}$ production proceeds via the excitation of the
$N^*$(1650), $N^*$(1710), $N^*$(1720) baryonic resonances. The 
parameters of the model at the nucleon-nucleon-meson vertices 
are determined by fitting the  
elastic nucleon-nucleon scattering with an effective interaction
based on the exchange of these four mesons, while those at the
resonance vertices are calculated from the known decay widths of
the resonances as well as the vector meson dominance model. Available
experimental data is described well by this approach.
The one-pion-exchange diagram dominates the production process at
both higher and lower beam energies. The $\rho$ and $\omega$ meson
exchanges make negligible contributions. However, the $\sigma$-exchange
processes contribute substantially to the total cross sections
at lower beam energies. The excitation of the  $N^*$(1710)
and $N^*$(1650) resonances dominate this reaction at beam momenta
above and below 3 GeV/c respectively. The interference among
the amplitudes of various resonance excitation processes is significant.
For beam energies very close to the $K^{+}$ production threshold the
hyperon-proton final state interaction effects are quite important.
The data is selective about the model used to describe the low energy
scattering of the two final state baryons. 
\end{abstract}
\addvspace{5mm}]


\vfill
\newpage
\section{Introduction}
In recent years there has been a considerable amount of interest
in the study of the associated production reaction $p$ + $p$ $\rightarrow$
$p$ + $K^{+}$ + $\Lambda$. This is expected to provide information on the
manifestation of quantum chromodynamics (QCD) in the non-perturbative
regime of energies larger than those of the low energy pion physics where 
the low energy theorem and partial conservation of axial current
(PCAC) constraints provide a useful insight into the relevant
physics~\cite{eric88}. The strangeness quantum number introduced
by this reaction leads to new degrees of freedom into this domain
which are expected to probe the admixture of $\bar{s}s$ quark pairs
in the nucleon wave function~\cite{albe96} and also the hyperon-nucleon and
hyperon-strange meson interactions~\cite{delo89,adel90}.

The elementary nucleon-nucleon-strange meson production cross sections
are the most important ingredients in the transport model studies of 
the $K^+$-meson production in the nucleus-nucleus collisions, which
provide information on not only the initial collision dynamics but
also the nuclear equation of state at high
density~\cite{mosel91,brown91,maru94,misk94,hart94,liko94,liko95,liko98}.
Furthermore, the enhancement in the strangeness production 
has been proposed as a signature for the formation of the
quark-gluon plasma in high energy nucleus-nucleus
collisions~\cite{rafe82,knol88}.

The experimental data on the $pp \rightarrow pK^{+}\Lambda$ reaction
is very scarce. The measurements performed in late 1960s and 1970s provide
total cross sections for this reaction at beam momenta larger than
2.80 GeV/c (see e.g.~\cite{land88}). With the advent of the high-duty
proton-synchrotron COSY at the Forschungszentrum, J\"ulich, it has
become possible to perform systematic studies of the associated strangeness
production at beam momenta very close to the reaction threshold
(2.340 GeV/c). The first round of experiments at COSY have already
added~\cite{juel98} 12 new data points to the data base. At near
threshold beam energies the final state interaction effects among the outgoing
particles are significant. Therefore, the new set of data are expected to
probe also the hyperon-nucleon and hyperon-strange meson interactions
along with the mechanism of the strangeness production in proton-proton
collisions.

The existing theoretical studies of this reaction are based either on
a single boson ($\pi$ or $K$ meson) exchange
mechanism~\cite{ferr60,yao62,wuko89,lage91} or a resonance
model~\cite{tsus97,fald97,sibi98}. In the first method, the $K^+$ 
production is assumed to take place essentially through the exchange of
one intermediate pion or $K$-meson; the excitation of any intermediate 
nucleon resonance is not considered. The $K$-meson exchange amplitudes
were found to dominate~\cite{ferr60,lage91} the production cross sections.
However, the relative sign of the pion and $K$-meson exchange amplitudes
was not fixed in this approach~\cite{lage91}. Furthermore, it has been
argued that the existing high energy data can be well reproduced considering
only the single pion-exchange process~\cite{yao62,wuko89} since the
contribution of the $K$-meson exchange amplitude can be compensated by
various parameters of the model.

In the resonance model~\cite{tsus94} of the strangeness production
in $pp$ collisions, the $\pi$, $\eta$, and $\rho$-meson exchanges are
included and the $K^+$ meson production proceeds via the excitation
of the $N^*$(1650), $N^*$(1710) and $N^*$(1720)
resonances~\cite{tsus97,sibi98}.
However, the terms in the total amplitude involving the interference
of various resonance contributions are neglected in these calculations.
Moreover, the parameters of the $NN\pi$ and $NN\rho$ vertices were taken from
the Bonn nucleon-nucleon potential which may not be adequate at higher
beam energies as these have been determined by fitting the $NN$
scattering data below the $NN\pi$ production threshold. At the same
time, the finite life-time of the $\rho$
meson has not been taken into account while calculating 
the relevant coupling constants from the experimental branching ratios.

In this paper, we investigate the associated $K^+$ production in the
proton-proton collisions in the framework of an effective Lagrangian
approach~\cite{wein68,pecc69,davi91,hagl91,benm95}, following and extending
our previous study~\cite{enge96,shya98} on $\pi^0$ and $\pi^+$  
production. Initial interaction between two incoming nucleons is 
modeled by an effective Lagrangian which is based on the exchange
of the $\pi$, $\rho$, $\omega$ and $\sigma$ mesons. The coupling
constants at the nucleon-nucleon-meson vertices are determined by 
directly fitting the T-matrices of the nucleon-nucleon ($NN$) scattering
in the relevant energy region~\cite{scha94}. The effective-Lagrangian
uses the pseudovector (PV) coupling for the nucleon-nucleon-pion
vertex (unlike the resonance model~\cite{tsus97}),
and thus incorporates the low energy theorems~\cite{domb73} of current
algebra and the hypothesis of partially conserved axial-vector current 
(PCAC). The $K^+$ production proceeds via excitation of the $N^*$(1650),
$N^*$(1710) and $N^*$(1720) intermediate baryonic resonance states which
have appreciable branching ratios for the decay into the $K^+\Lambda$
channel. The interference terms between various intermediate resonance 
states are included which marks a major difference between our work and the
resonance model~\cite{tsus97}. To describe the recent near threshold
data, the final state interaction between the outgoing
particles is included within the framework of the Watson-Migdal
theory~\cite{shya98}.

The remainder of this paper is organized in the following way. Section II
contains details of our theoretical approach. Section III comprises the
results of our analysis and their critical discussion. The summary and
conclusions of our work are presented in section IV. Finally some
technical details are given in appendix A.

\section{FORMALISM}

We consider the tree-level structure (Fig. 1) of the amplitudes for
the associated $K^+\Lambda$ production in proton-proton collisions, which   
proceeds via the excitation of the $N^*$(1650)($\frac{1}{2}^{-}$),
$N^*$(1710)($\frac{1}{2}^{+}$) and $N^*$(1720)($\frac{3}{2}^{+}$) 
intermediate resonances. To evaluate these amplitudes within the
effective Lagrangian approach, one needs to know the effective Lagrangians
(and the coupling constants appearing therein) at (i) nucleon-nucleon-meson,
(ii) resonance-nucleon-meson, and (iii) resonance-$K^+$-hyperon vertices.
These are discussed in the following subsections.
 
\subsection{Nucleon-nucleon-meson vertex}

The parameters for $NN$ vertices are determined by fitting the $NN$
elastic scattering T matrix with an effective $NN$ interaction based
on the $\pi$, $\rho$, $\omega$ and $\sigma$ meson exchanges. The
effective meson-$NN$ Lagrangians are
  
\begin{figure}[here]
\begin{center}
\epsfxsize=7.00cm
\epsfbox{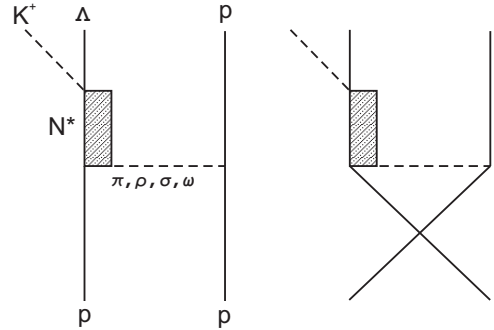}
\end{center}
\vskip -4truecm
\caption[C1]{
Feynman diagrams for $K^{+}\Lambda$ production in $pp$
collisions. The diagram on the left shows the direct process while that
on the right the exchange one.}
\label{fig:figa}
\end{figure}

\begin{eqnarray}
{\cal L}_{NN\pi} & = & -\frac{g_{NN\pi}}{2m_N} {\bar{\Psi}}_N \gamma _5
                             {\gamma}_{\mu} \tauiso
                            \cdot (\partial ^\mu {\bf \Phi}_\pi) \Psi _N. \\
{\cal L}_{NN\rho} &=&- g_{NN\rho} \bar{\Psi}_N \left( \gmu + \frac{k_\rho}
                   {2 m_N} \sigma_{\mu\nu} \partial^\nu\right)\nonumber \\
                & & \times \tauiso \cdot \mbox{\boldmath $\rho$}^\mu \Psi_N. \\
{\cal L}_{NN\omega} &=&- g_{NN\omega} \bar{\Psi}_N \left( \gmu + \frac{k_\omega}
                     {2 m_N} \sigma_{\mu\nu} \partial^\nu\right)
                          \omega^\mu \Psi_N.   \\
{\cal L}_{NN\sigma} &=& g_{NN\sigma} \bar{\Psi}_N \sigma \Psi_N.
\end{eqnarray}
In Eqs. (2) - (3) $\sigma_{\mu\nu}$ is defined as 
\begin{eqnarray}
\sigma_{\mu\nu} & = & { i \over 2}(\gamma_\mu\gamma_\nu - \gamma_\nu\gamma_\mu)
\end{eqnarray}
We have used the notations and conventions of Bjorken and Drell~\cite{bjor64}. 
In Eq. (1) $m_N$ denotes the nucleon mass. It should be noted that we have used
a PV coupling for the $NN\pi$ vertex.
Since we use these Lagrangians to directly model the T-matrix, we have
also included a nucleon-nucleon-axial-vector-isovector vertex, with the
effective Lagrangian given by
\begin{eqnarray}
{\cal L}_{NNA} & = & g_{NNA} {\bar {\Psi}} \gamma_5 \gamma_\mu \tauiso \Psi
                     \cdot {\bf {A}}^\mu,
\end{eqnarray}
where $A$ represents the axial-vector meson field. This term is introduced
because in the limit of large axial meson mass ($m_A$) it cures the 
unphysical behavior in the angular distribution of $NN$ scattering caused by
the contact term in the one-pion exchange amplitude~\cite{scha94}, if
$g_{NNA}$ is chosen to be
\begin{eqnarray}
g_{NNA} =  \frac{1}{\sqrt{3}} m_A \left(\frac{f_\pi}{m_\pi}\right).
\end{eqnarray}
with very large ($\gg m_N$) $m_A$. $f_\pi$ appearing in Eq. (7) is 
related to $g_{NN\pi}$ as $f_\pi = (\frac{g_{NN\pi}}{2m_N})m_\pi$.

It should be noted that the contact term of the coordinate space potential
corresponding to one pion exchange term, is effectively switched off by 
the repulsive hard core of the nucleon-nucleon interaction. However,
in the effective Lagrangian description this term has to be explicitly
subtracted in order to avoid the unphysical behavior of the $NN$ elastic 
scattering cross sections. This is achieved by the inclusion of a term
corresponding to the exchange of a axial-vector-isovector meson as 
described above.

We introduce, at each interaction vertex, the form factor
\begin{eqnarray}
F_{i}^{NN} & = & \left (\frac{\lambda_i^{2} - m_i^{2}}{\lambda_i^{2} - q_i^{2}}
        \right ), i= \pi, \rho, \sigma, \omega,
\end{eqnarray}
where $q_i$ and $m_i$ are the four momentum and mass of the $i$th 
exchanged meson respectively. The form factors suppress the contributions of
high momenta and the parameter $\lambda_i$, which governs the
range of suppression, can be related to the hadron size. Since
$NN$ elastic scattering cross sections decrease gradually with the beam
energy (beyond certain value), we take energy dependent meson-nucleon
coupling constants of the following form 
\begin{eqnarray}
g(\sqrt{s}) & = & g_{0} exp(-\ell \sqrt{s}),
\end{eqnarray}
in order to reproduce these data in the entire range of beam energies. The
parameters, $g_0$, $\lambda$ and $\ell$ were determined by fitting
to the elastic proton-proton and proton-neutron scattering data at the
beam energies in the range of 400 MeV to 4.0 GeV~\cite{enge96,scha94}.
It may be noted that this procedure also fixes the sign of the
effective Lagrangians (Eq. (1)-(4) and (6)).
The values of various parameters are shown in Table 1, which are the 
same as those used in the calculations of the pion production in proton-proton
collisions~\cite{enge96,shya98}.  Thus we ensure that the $NN$ 
elastic scattering channel remains the same in the description of various
inelastic channels within this approach, as it should be. 
\vbox{  
\begin{table}[here]
\begin{center}
\caption  {Coupling constants for the $NN$-meson
vertices used in the calculations}
\vspace{0.5cm}
\begin{tabular}{|c|c|c|c|c|} \hline
 Meson & $g^2/4\pi$ & $\ell$ & $\lambda$ & mass \\
       &             &        & (\footnotesize{GeV} ) & (\footnotesize{GeV})
 \\ \hline
$\pi    $ & 12.562 & 0.1133 & 1.005 & 0.138 \\
$\sigma $ & 2.340  & 0.1070 & 1.952 & 0.550 \\
$\omega $ & 46.035 & 0.0985 & 0.984 & 0.783 \\
$\rho   $ & 0.317  & 0.1800 & 1.607 & 0.770 \\
$k_{\rho}$ = 6.033, $k_{\omega}$ = 0.0 & & & & \\ \hline
\end{tabular} 
\end{center}
\end{table}
}

\subsection{Resonance-nucleon-meson vertex}

As the $\Lambda$-hyperon has zero isospin, only isospin-1/2 nucleon
resonances are allowed. Below 2 GeV center of mass (c.m.)
energy, only three resonances,
$N^*$(1650), $N^*$(1710), and $N^*$(1720),
have significant decay branching ratios (3-11$\%$, 5-25$\%$ and 1-15$\%$
respectively~\cite{part98}) into the $K^{+}\Lambda$ channel.
In this work only these three resonances have been considered.
The $N^*$(1700) resonance having very small (and uncertain) branching
ratio for the decay to this channel has been excluded. 

Since all the three resonances can couple to the meson-nucleon channel 
considered in the previous section, we require the effective Lagrangians
for all the four resonance-nucleon-meson vertices corresponding to
all the included resonances. For the coupling of the spin-1/2- resonances
to pion we again have the choice of PS or PV couplings. The corresponding
effective Lagrangians can be written as~\cite{benm95,feus97}
\begin{eqnarray}
{\cal L}^{PV}_{N_{1/2}^*N\pi} & = & -\frac{g_{N_{1/2}^*N\pi}}{M}
                          {\bar{\Psi}}_{N^*} {\Gamma}_{\mu} \tauiso
                       \cdot (\partial ^\mu {\bf \Phi}_\pi)
                        \Psi _N \nonumber \\
                      & &    + {\rm h.c.}\\ 
{\cal L}^{PS}_{N_{1/2}^*N\pi} & = & -g_{N_{1/2}^*N\pi}
                          {\bar{\Psi}}_{N^*} i{\Gamma} \tauiso
                        {\bf \Phi}_\pi \Psi _N 
                          + {\rm h.c.}, 
\end{eqnarray}
where $M \,=\,(m_{N^*}\,\pm\,m_N)$, with the upper sign for even parity and
lower sign for odd parity resonance. The operators $\Gamma$, $\Gamma_\mu$,
are given by,
\begin{eqnarray}
\Gamma = \gamma_5,\,\, \Gamma_\mu = \gamma_5 \gamma_\mu,\\
\Gamma = 1, \,\, \Gamma_\mu = \gamma_\mu,
\end{eqnarray}
for resonances of even and odd parities, respectively. 
We have performed calculations with both of these couplings. The effective
Lagrangians for the coupling of resonances to other mesons are
similar to those given by Eq. (2)-(4),
\begin{eqnarray}
{\cal L}_{N_{1/2}^*N\rho} & = &- g_{N_{1/2}^*N\rho} \bar{\Psi}_{N^*}
                           \frac{1} {2m_N}
                            \Gamma_{\mu\nu} \partial^\nu
                      \tauiso \cdot \mbox{\boldmath $\rho$}^\mu \Psi_N
                       \nonumber \\
                          &   & + {\rm h.c.}\\
{\cal L}_{N_{1/2}^*N\omega} &=&- g_{N_{1/2}^*N\omega} \bar{\Psi}_{N^*}
                         \frac{1}{2m_N}
                         \Gamma_{\mu\nu} \partial^\nu
                          \omega^\mu \Psi_N \nonumber \\
                        &   &  + {\rm h.c.}  \\
{\cal L}_{N_{1/2}^*N\sigma} & = & g_{N_{1/2}^*N\sigma} \bar{\Psi}_{N^*}
                             \Gamma^\prime\sigma \Psi_N \nonumber \\
                        &   &      + {\rm h.c.},
\end{eqnarray}
The operators ${\Gamma^\prime}$ and $\Gamma_{\mu\nu}$ are,
\begin{eqnarray}
\Gamma^\prime = 1, \,\, \Gamma_{\mu\nu} = \sigma_{\mu\nu}\\
\Gamma^\prime = \gamma_5,\, \, \Gamma_{\mu\nu} = \gamma_5 \sigma_{\mu\nu}
\end{eqnarray}
for resonances of even and odd parities, respectively.

The even parity isospin-1/2 $N^*$(1720) resonance is a spin-3/2 nucleon
excited state. We have used the following effective Lagrangians for
vertices involving this resonance~\cite{benm95,feus97}
\begin{eqnarray}
{\cal L}_{N^*N\pi} & = & \frac{g_{N_{3/2}^*N\pi}}{m_\pi}
                         {\bar{\Psi}}_{\mu}
                         {\tauiso} \cdot \partial ^{\mu}
                         {\bf \Phi}_\pi \Psi _N 
                         + {\rm h.c.}, \\
{\cal L}_{N^*N\rho} & = & {\rm i} \frac{g_{N_{3/2}^*N\rho}}{m_{N^*} + m_N}
                        {\bar{\Psi}}_{\mu}\tauiso \left(
                        \partial^\nu \mbox{\boldmath $\rho$}^\mu -
                        \partial^\mu \mbox{\boldmath $\rho$}^\nu
                        \right) \nonumber \\
                    &   &    \gamma_\nu \gf\Psi _N + {\rm h.c.},\\
{\cal L}_{N^*N\omega} & = & {\rm i} \frac{g_{N_{3/2}^*N\omega}}{m_{N^*} + m_N}
                        {\bar{\Psi}}_{\mu} \tauiso \left(
                        \partial^\nu \omega^\mu -
                        \partial^\mu \omega^\nu \right) \nonumber \\
                    &   &     \gamma_\nu \gf\Psi _N + {\rm h.c.},\\
{\cal L}_{N^*N\sigma} & = & \frac{g_{N_{3/2}^*N\sigma}}{m_\sigma}
                         {\bar{\Psi}}_{\mu}
                         {\tauiso} \cdot (\partial ^{\mu}
                         {\sigma}) \Psi _N 
                         + {\rm h.c.}. 
\end{eqnarray}
Here, ${\bar {\Psi}}_\mu$ is the $N^*$(1720) vector spinor.
It should be remarked here that an operator
$\Theta_{\alpha\mu}(z) = g_{\alpha\nu} -\frac{1}{2}(1+2z)\gamma_\alpha
\gamma_\mu$ has also been included in
the vector spinor vertex in Refs.~\cite{benm95,feus97}. This operator
describes the off-shell admixture of the spin-1/2 fields~\cite{benm89}.
The choice of the off-shell parameter $z$ is arbitrary and it is
treated as a free parameter to be determined by fitting to the data.
This operator can be easily introduced in Eqs. (19)-(22) which will
introduce four additional free parameters in our calculations. We however,
work with the Lagrangians as given in Eqs. (19)-(22), which are
identical to those given in~\cite{benm95,feus97} for $z$ = 0.5. 
  
\subsection{Resonance-hyperon-strange meson vertex}

For vertices involving spin-1/2 resonances, there is again the PS and
PV coupling option. In principle one can select a linear combination
of both and fit the PS/PV ratio to the data. However, to minimize the
number of parameters we choose either PS or PV coupling at a time. 
The effective Lagrangians for the $N^*$$\Lambda$$K^+$ vertex is 
written in the following way~\cite{feus97},

For spin-1/2 resonance,
\begin{eqnarray}
{\cal L}^{PV}_{N_{1/2}^*\Lambda K^+} & = &
                    -\frac{g_{N^*_{1/2}\Lambda K^+}}{M^\prime}
                          {\bar{\Psi}}_{N^*} {\Gamma}_{\mu} \tauiso
                       \cdot (\partial ^\mu {\bf \Phi}_{K^{+}})
                        \Psi _N \nonumber \\
                     &  &   + {\rm h.c.},\\ 
{\cal L}^{PS}_{N_{1/2}^*\Lambda K^+} & = & -g_{N^*_{1/2}\Lambda K^+}
                          {\bar{\Psi}}_{N^*} {i\Gamma} \tauiso
                        {\bf \Phi}_{K^{+}} \Psi _N \nonumber \\
                     &  &   + {\rm h.c.}, 
\end{eqnarray}
where $M^\prime = m_{N^*} \pm m_{\Lambda}$, with the upper sign for even parity
and lower sign for odd parity resonance. 

For spin-3/2 resonance, 
\begin{eqnarray}
{\cal L}_{N_{3/2}^*\Lambda K^+} & = & \frac{g_{N^*_{3/2}\Lambda K^+}}{m_{K^+}}
                         {\bar{\Psi}}_{\mu}
                         {\tauiso} \cdot \partial ^{\mu}
                         {\bf \Phi}_{K^{+}} \Psi _N \nonumber \\
                    &  &     + {\rm h.c.}. 
\end{eqnarray}

\subsection{Coupling constants for resonances}

The resonance couplings are determined from the experimentally 
observed quantities such as the branching ratios for the decay of the
resonances to the corresponding channels. 

The partial width for the decay of a resonance (in its rest frame) of
mass $M_{N^*}$ into a meson of mass $m_m$ and energy $E_m$ and a nucleon  
is written in terms of the Lorenz invariant matrix element
$\cal{M}$ as
\begin{eqnarray}
d\Gamma & = & \frac{(2\pi)^4}{2M_{N^*}} |{\cal{M}}|^2
                 \delta^4(P_{N^*}-p_m-p_N)
            \frac{d^3p_m}{(2\pi)^3 2E_m} \nonumber \\
        &   & \times    \frac{m_N}{E_N}
             \frac{d^3p_N}{(2\pi)^3},
\end{eqnarray}
In case of the meson (in the decay channel) having a finite life time
for the decay to another channel (e.g $\rho$ $\rightarrow$ $\pi\pi$ ),
an integration over the phase-space for this decay must be
included~\cite{frim97,rapp97,pete98}.
 
\vskip .1in

{\bf (i). $N^*N\pi$ vertex}

\vskip .1in

For the spin-1/2 resonance the $N^*N\pi$ decay width, with the PS coupling, 
is given by
\begin{eqnarray}
\Gamma_{N_{1/2}^*N\pi} & = & \frac{3}{4\pi}g_{N_{1/2}^*N\pi}^2
            \frac{E_N \pm m_N}{m_{N^*}} p_\pi^{cm},
\end{eqnarray}
while that with the corresponding PV coupling is
\begin{eqnarray}
\Gamma_{N_{1/2}^*N\pi} & = & \frac{3}{4\pi}
                \left(\frac{g_{N_{1/2}^*N\pi}}{M}\right)^2 \nonumber \\ 
 &  & \times \left[\frac{2E_\pi(E_NE_\pi + (p_\pi^{cm})^2)
                 - m_\pi^2(E_N \pm m_N)} {m_{N^*}}\right] \nonumber \\
  & & \times       p_\pi^{cm}, 
\end{eqnarray}
where
\begin{eqnarray}
p_\pi^{cm} & = & \frac{[m_{N^*}^2 - (m_N + m_\pi)^2]}
                       {4m_{N^*}^2} \nonumber \\
                 & & \times      [m_{N^*}^2 - (m_N-m_\pi)^2], \\
E_N & = & \sqrt{(p_\pi^{cm})^2 + m_N^2},\\
E_\pi & = & \sqrt{(p_\pi^{cm})^2 + m_\pi^2}.
\end{eqnarray}

For spin-3/2 resonance the $N^*N\pi$ decay width is
\begin{eqnarray}
\Gamma_{N_{3/2}^*N\pi} & = & \frac{1}{12\pi}\left(\frac{g_{N_{3/2}^*N\pi}}
                             {m_\pi}\right)^2 
                        \frac{E_N \pm m_N}{m_N^*} (p_\pi^{cm})^3.
\end{eqnarray}
The plus and minus sign in Eqs (27) corresponds to odd and even parity
resonances respectively, while in Eqs. (28) and (32) reverse is the case.

\vskip .1in

{\bf (ii). $N^*N\rho$ vertex}

\vskip .1in

The partial decay width of each resonance for the decay into nucleon
and two pions via the $\rho$ meson is given by
\begin{eqnarray}
\Gamma(m_{N^*}) & = & 2 \int_{2m_\pi}^{m_{N^*}-m_N}
                    dm \;m \;\Gamma^*(m) S(m).
\end{eqnarray}
In this equation the spectral function $S(m)$ is defined as, 
\begin{eqnarray}
S(m) & = & -\frac{1}{\pi} \;Im \;D_\rho(m),
\end{eqnarray}
where 
\begin{eqnarray}
D_\rho(m) & = & \frac{1}{m^2-m_\rho^2+im \Gamma_{\rho\rightarrow\pi\pi}},
\end{eqnarray}
with
\begin{eqnarray}
\Gamma_{\rho \rightarrow \pi\pi} & = & \Gamma^0_{\rho \rightarrow \pi\pi}
                               \frac{m_\rho^2}{m^2} 
                \left[\frac{p_{\rho\pi\pi}(m)} {p_{\rho\pi\pi}(m_\rho)}
                 \right]^3.
\end{eqnarray}
The value of $\Gamma^0_{\rho \rightarrow \pi\pi}$ is taken to be 150 MeV. 
The $\rho \rightarrow \pi\pi$ decay four-momenta $p_{\rho\pi\pi}$
are 
\begin{eqnarray}
p_{\rho\pi\pi}(m) & = & \frac{[m^2 - 4m_\pi^2 ][m^2 ]}
                       {4m^2}. 
\end{eqnarray}
In Eq. (33) $\Gamma^*(m)$ is defined in the following way,
\vskip .1in
For spin-1/2 even parity resonance
\begin{eqnarray}
\Gamma^*(m) & = & \frac{1}{4\pi}\left( g_{N_{1/2}^*N\rho} \over {2m_N}
                \right )^2 \nonumber \\
            &   & \times   \left[\frac{4(E_N^*+E_m)(p_m^{cm})^2 +
                 3(E_N^*-m_N)m^2}{m_{N^*}}\right ] \nonumber \\
            &   & \times       p_\pi^{cm},\\ 
E_N^* & = & \sqrt{(p_m^{cm})^2 + m_N^2},\\
E_m & = & \sqrt{(p_m^{cm})^2 + m^2},
\end{eqnarray}
where $p_m^{cm}$ is given in the same way as Eq. (29) with $m_\pi$ 
replaced by $m$.
\vskip .1in
For spin-1/2 odd parity resonance
\begin{eqnarray}
\Gamma^*(m) & = & \frac{1}{4\pi}\left( g_{N_{1/2}^*N\rho} \over {2m_N}
                \right )^2  \nonumber \\
            &   & \times  \left[\frac{-4(E_N^*+E_m)(p_m^{cm})^2 -
                 3(E_N^*+m_N)m^2}{m_{N^*}}\right ] \nonumber \\ 
  & & \times       p_\pi^{cm}. 
\end{eqnarray}
\vskip .1in
For spin-3/2 even parity resonance
\begin{eqnarray}
\Gamma^*(m) & = & \frac{1}{12\pi}\left( g_{N_{3/2}^*N\rho} \over {m_{N^*}+m_N}
                \right )^2 \nonumber \\
            & &  \times  \left[\frac{2(2E_N^*+E_m)(p_m^{cm})^2 +
                 3(E_N^*-m_N)m^2}{m_{N^*}}\right ] \nonumber \\
  & & \times       p_\pi^{cm}. 
\end{eqnarray}

\vskip .1in

{\bf (iii). $N^*N\omega$ vertex}

\vskip .1in

Since the resonances considered in this study have no known branching
ratios for the decay into the $N\omega$ channel, we determine the
coupling constants for the $N^*N\omega$ vertices by the strict
vector meson dominance (VMD) hypothesis~\cite{saku69}, which  
is based essentially on the assumption that the coupling of photons on hadrons 
takes place through a vector meson.

The $N^*N\gamma$ partial widths are given as following: 

\begin{flushleft}For spin-1/2 even parity resonance,\end{flushleft}
\begin{eqnarray}
\Gamma_{N^*N\gamma} & = & {1 \over \pi} {m_N \over m_{N^*}} (\mu_{N^*})^2
                          (q_f^3)
\end{eqnarray}
For spin-1/2 odd parity resonance,
\begin{eqnarray}
\Gamma_{N^*N\gamma} & = & {3 \over 2} {m_N \over m_{N^*}} (\mu_{N^*})^2
                          (m_N^2+{2 \over 3}q_f^2)q_f
\end{eqnarray}
For spin-3/2 even parity resonance,
\begin{eqnarray}
\Gamma_{N^*N\gamma} & = & {1 \over \pi} {m_N \over m_{N^*}} (\mu_{N^*})^2
                          (q_f^3)
\end{eqnarray}
In these equations, $q_f$ $=$ $((m_{N^*}^2 - m_N^2)/2m_{N^*})$. The value of
$\mu_{N^*}$ is determined by fitting to the $N\gamma$ partial width of
each resonance which is given in terms of the helicity amplitudes 
$A_{1/2}$ and $A_{3/2}$ by~\cite{part98}
\begin{eqnarray}
\Gamma_\gamma & = & {q_f^2 \over \pi} {{2m_N} \over {(2J+1)m_{N^*}}}
                 \left[ |A_{1/2}|^2 + |A_{3/2}|^2 \right ],
\end{eqnarray}
where $J$ is the resonance spin. $\mu_{N^*}$ is written as the ratio of the 
couplings at $N^*\omega$ and $\omega\gamma$ vertices as
\begin{eqnarray}
\mu_{N^*} & = & e {g_{N^*\omega}} \over {g_{\omega\gamma}}
\end{eqnarray} 
Using above equations together with the experimental helicity amplitudes
the values of the coupling constants for the $N^*N\omega$ vertices can 
be determined. We have used $g_{\omega \gamma}$ = 17 in our calculations.
\vskip .1in

{\bf (iv). $N^*N\sigma$ vertex}

\vskip .1in

As the sigma meson is, most of the time, a resonance of two
pions~\cite{bonn87} in the S-state, the coupling constants for the
$N^*N\sigma$ vertices are determined from the branching ratios of
the decay of the resonances into $N(\pi \pi)^{\ell=0}$. We, however,
reduce the experimental values of these ratios by two third to account
for the fact that this correlated state provides only about 2/3 of the
total 2$\pi$-exchange.
The expressions for the partial widths are similar to those given by
Eqs. (27)-(32). 
\vskip .1in

{\bf (iv). $N^*\Lambda K^+$ vertex}

\vskip .1in

The coupling constants for the $N^*\Lambda K^+$ vertex are determined
from the experimental branching ratios for the $N^*$ $\rightarrow$ $\Lambda$
$K^+$ decay. The expressions for the decay widths are similar to those  
given by Eqs. (27) - (32).

We assume that the off-shell dependence of the $NN^*$ vertices are determined
solely by multiplying the vertex constants by the form factors, 
which have the dipole form~\cite{enge96,koen81}
\begin{eqnarray}
F_{i}^{NN^*} & = & \left (\frac{(\lambda_i^{N^*})^2 - m_i^{2}}
                   {(\lambda_i^{N^*})^2 - q_i^{2}}
                    \right )^2, i= \pi, \rho, \sigma, \omega,
\end{eqnarray}

The resonance properties used in the calculations of the decay widths are
given in Table 2, where the resulting coupling constants and the adopted
values of the cut-off parameters ($\lambda_i^{NN^*}$) are also given.
It may be noted that we have fixed the latter to one value in order to
minimize the number of free parameters.

It should however, be stressed that the branching ratios determine only
the square of the corresponding coupling constants; thus their signs remain 
uncertain in this method. Predictions from independent calculations 
(${\it e.g}$ the quark model) can, however, be used to constrain these
signs. The magnitude as well signs of the coupling constants for the
$N^*N\pi$, $N^*\Lambda K$, $N^*N\rho$, and $N^*N(\pi \pi)_{s-wave}$
vertices were determined by Feuster and Mosel~\cite{feus98} and Manley 
and Saleski~\cite{manl92} in their analysis of the pion-nucleon data
involving the final states $\pi N$, $\pi \pi N$, $\eta N$, and $K\Lambda$.
Predictions for some of these quantities are also given in the 
constituent quark model calculations of Capstick and Roberts~\cite{caps94}.
Guided by the results of these studies, we have chosen the positive
sign for the coupling constants for these vertices. Unfortunately,
the quark model calculations for the $N^*N\omega$ vertices are still
sparse and an unambiguous prediction for the signs of the corresponding
coupling constants may not be possible at this stage~\cite{stan93}.
Nevertheless, we have chosen a positive sign for the coupling constants
for these vertices as well.
 
\vbox{
\begin{table}[here]
\begin{center}
\caption  {Coupling constants and cut-off parameters for the  
$N^*N$-meson and $N^*$-hyperon-meson vertices used in the calculations}
\vspace{0.5cm}
\begin{tabular}{|c|c|c|c|c|c|} \hline
Resonance &Width &Decay &Adopted & $g^2/4\pi$ & cut-off \\
        &(\footnotesize{GeV})& channel& value of the& & (\footnotesize{GeV}) \\ 
        &     &      &  branching &       & \\
        &     &      &  ratio     &           & \\    \hline
$N^*$(1710)&0.100 & $N\pi$   &0.150     & 0.0863 & 0.850 \\
           &    &$N\rho$   &0.150     & 1.3653 & 0.850 \\
           &    &$N\omega$ &          & 0.1189 & 0.850 \\
           &    &$N\sigma$ &0.170     & 0.0361 & 0.850 \\
           &    &$\Lambda K$ &0.150   & 2.9761 & 0.850 \\ 
$N^*$(1720)&0.150 &$N\pi$    &0.100     & 0.0023 & 0.850 \\
           &    &$N\rho$   &0.700     & 90.637 & 0.850 \\
           &    &$N\omega$ &          & 22.810 & 0.850 \\
           &    &$N\sigma$ &0.120     & 0.1926 & 0.850 \\
           &    &$\Lambda K$ &0.080   & 0.0817 & 0.850 \\ 
$N^*$(1650)&0.150 &$N\pi$    &0.700     & 0.0521 & 0.850 \\
           &    &$N\rho$    &0.08     & 0.5447 & 0.850 \\
           &    &$N\omega$ &          & 0.2582 & 0.850 \\
           &    &$N\sigma$ &0.025     & 0.2882 & 0.850 \\
           &    &$\Lambda K$ &0.070   & 0.0485 & 0.850 \\
\hline 
\end{tabular}
\end{center}
\end{table}
}

\subsection{Propagators}

We require the propagators for various mesons and nucleon resonances in the
calculation of the amplitudes. The propagators for pion, $\rho$ meson and
axial-vector mesons are given by
 
\begin{eqnarray}
G_\pi(q) & = & {i \over (q^2 - m_\pi^2)}\\
G_\rho^{\mu\nu}(q) & = & -i\left({g^{\mu\nu}-q^\mu q^\nu/q^2}
                          \over {q^2 - m_\rho^2} \right)\\
G_A^{\mu\nu}(q) & = & -i\left(\frac{g^{\mu\nu}}{q^2-m_A^2}\right)
\end{eqnarray}
In Eq. (51), the mass of the axial meson is taken to be very large (188 GeV),
as the corresponding amplitude is that of the contact term. The propagators
for $\omega$ and $\sigma$ mesons are similar to those given
by Eqs. (50) and (49) respectively.

The propagators for spin-1/2 and spin-3/2 resonances are 
\begin{eqnarray}
G_{N^*_{1/2}} (p) & = & i\left( {p_\eta\gamma^\eta + m_{N^*_{1/2}}} \over
                       {p^2 - (m_{N^*_{1/2}}-i(\Gamma_{N^*_{1/2}}/2))^2}
                         \right ),\\
G_{N^*_{3/2}}^{\mu \nu} (p) & = & -\frac{i(p\!\!\!/ + m_{N^*_{3/2}})}
                    {p^2 - (m_{N^*_{3/2}}-i(\Gamma_{N^*_{3/2}}/2))^2}
                               \nonumber \\
                 &   &\times [g^{\mu \nu} - \frac{1}{3}\gamma^\mu \gamma^\nu
                              - \frac{2}{3m_{N^*_{3/2}}^2} p^\mu p^\nu
                               \nonumber \\
                 &   &    + \frac{1}{3m_{N^*_{3/2}}^2}
                                ( p^\mu \gamma^\nu - p^\nu \gamma^\mu )].
\end{eqnarray}
In Eqs. (52) - (53), $\Gamma_{N^*}$ is the total width of the resonance which
is introduced in the denominator term $(p^2-m_{N^*}^2)$ to account for the
fact that the resonances are not the stable particles; they
have a finite life time for the decay into various channels. $\Gamma_{N^*}$ 
is a function of the center of mass momentum of the decay channel, and it is
taken to be the sum of the widths for pion and rho decay (the other decay  
channels are considered only implicitly by adding their branching ratios
to that of the pion channel)
\begin{eqnarray}
\Gamma_{N^*} & = & \Gamma_{N^*\rightarrow N\pi} +
                      \Gamma_{N^*\rightarrow N\rho}
\end{eqnarray}
$\Gamma_{N^*\rightarrow N\rho}$ is calculated according to Eq. (33). 
$\Gamma_{N^*\rightarrow N\pi}$ is taken to be
\begin{eqnarray}
\Gamma_{N^*\rightarrow N\pi} & = & \Gamma_0 \left({p_{\pi R}^{cm}} \over 
                                 {p_\pi^{cm}}\right)^{2\ell+1},
\end{eqnarray}
where $\ell$ is the orbital angular momentum of the resonance. $p_\pi^{cm}$
is as defined in Eq. (29) and $p_{\pi R}^{cm}$ is given by the same equation
with $m_{N^*}$ replaced by $p$ of Eqs. (52) - (53). $\Gamma_0$ is 
taken to be the total on-shell width of the resonance minus the
corresponding width for the nucleon-$\rho$ meson decay channel.

\subsection{Amplitudes and cross sections}

After having established the effective Lagrangians, coupling constants and
form of the propagators, we can now proceed to calculate the amplitudes
for various diagrams associated with the $pp$ $\rightarrow$ $p\Lambda K^+$
reaction. These amplitudes can be written by following the well known 
Feynman rules~\cite{itzy88} and calculated numerically by following
${\it e.g.}$ the techniques discussed in~\cite{enge96}. The isospin
part is treated separately. This 
gives rise to a constant factor for each graph, which is unity for the
reaction under study. It should be noted that the signs of 
various amplitudes are fixed by those of the effective Lagrangian 
densities, coupling constants and propagators 
as described above. These signs are not allowed to change anywhere in
the calculations. 

The general formula for the invariant cross section of the $p$ + $p$
$\rightarrow$ $p$ + $\Lambda$ + $K^+$ reaction is written as 
\begin{eqnarray}
d\sigma & = & \frac{m_N^3m_\Lambda}{2\sqrt{[(p_1 \cdot p_2)^2-m_N^4]}}
                     \frac{1}{(2\pi)^5}\delta^4(P_f-P_i)|A_{fi}|^2 \nonumber \\
                 & & \times    \prod_{a=1}^3 \frac{d^3p_a}{E_a},
\end{eqnarray}
where $A_{fi}$ represents the total amplitude, $P_i$ and $P_f$
the sum of all the momenta in the initial and final states, respectively, and
$p_a$ the momenta of the three particles in the final state. The
corresponding cross sections in the laboratory or center of mass systems
can be written from this equation by imposing the relevant conditions.

\subsection{Final state interaction}

For describing the data for the $pp$ $\rightarrow$ $p\Lambda K^+$,
reaction at beam energies very close to the threshold, consideration 
of the final state interaction (FSI) among the three out going
particles is important. As there exists no theory of the FSI 
effects in the presence of three strongly interacting particles, we
follow here an approximate scheme in line exactly with Watson-Migdal
theory of  FSI~\cite{wats52}. In this approach the energy dependence of
the cross section due to FSI is separated from that of the primary 
production amplitude. This method has been applied earlier to 
study the low momentum behavior of the $\eta$ meson~\cite{moal96}
and pion spectra~\cite{duba86,shya98,meis98} measured in proton-proton
collisions. We write for the total amplitude
\begin{eqnarray}
A_{fi} & = & M_{fi}(pp \rightarrow p\Lambda K^+) \cdot T_{ff},
\end{eqnarray}
where $M_{fi}(pp \rightarrow p\Lambda K^+)$ is the primary production
amplitude as discussed above, while $T_{ff}$ describes the re-scattering 
among the final particles which goes to unity in the limit of no FSI. 
The latter is taken to be the coherent sum of the two-body on-mass-shell
elastic scattering amplitudes of the particles involved in the final 
channel. 
\begin{eqnarray}
T_{ff} & = & \sum_{i=1}^3 t_i^{\ell_i}(q_i), 
\end{eqnarray}
where $t_i$ represents the two-body on-shell elastic scattering amplitude
(of the interacting particles pair $j-k$) in the three-body space
with the $i$th particle being the spectator. $\ell_i$ and $q_i$ denote the 
partial wave and relative momentum of the $j-k$ particle pair respectively.

An assumption inherent in the approximation given by Eq. (65) is that
the reaction takes place over a small region of space, a condition 
fulfilled rather well in near threshold reactions involving heavy mesons.
This allows us to
express the amplitudes $t_i$ in terms of the inverse of the Jost function,
$J_{\ell_i}(q_i)$~\cite{wats69,shya98}. In the analysis presented in this
paper we assume $\ell_i\;=\;0$, for all the three pairs $j-k$ and  
use the modified Cini-Fubini-Stanghellini formula~\cite{noye72} for the 
effective range expansion of the phase-shift ($\delta_{0i}$) of the
relevant pair
\begin{eqnarray}
C_0^2 \;q_i cot\delta_{0i} + 2 q_i\eta\; h(\eta) = (1/a_i) +
                                         (1/2) r_{0i}q_i^2, 
\end{eqnarray}
to calculate the corresponding Jost function. It may be noted 
that in case of the pair $j-k$ involving uncharged particle(s), 
the second terms on the left hand side of Eq. (67) vanishes and
$C_0^2$ goes to unity. In this equation 
$r_{0i}$ and $a_{i}$ are the effective range and 
scattering length parameters respectively for the $j-k$ interacting pair.
$\eta$ is the corresponding Coulomb parameter and
\begin{eqnarray}
C_0^2 & = & \frac{2\pi\eta}{e^{2\pi\eta}-1},\\
h(\eta) & = & \sum_{n=1}^\infty \frac{\eta^2}{n(n^2+\eta^2)} - 0.5772 
                    -ln(\eta).
\end{eqnarray}                   
In this case we have
\begin{eqnarray}
t_i^0(q_i)  =  (J_0(q_i))^{-1} = \frac{(q_i^2+\alpha_i^2)r_{0i}^c/2}
                           {1/a_{i}^c+(r_{0i}^c/2)q_i^2-iq_i},
\end{eqnarray}
where $\alpha$ is given by
\begin{eqnarray}
\alpha & = & (1/r_{0i}^c)[1+(1+2r_{0i}^c/a_{i}^c)^{1/2}],
\end{eqnarray} 
and $a_i^c$ and $r^c_{0i}$ are defined as
\begin{eqnarray}
\frac{1}{a_i^c} & = & \frac{1}{C_0^2}(\frac{1}{a_i}-2q_i\eta\; h(\eta))\\
r_{0i}^c & = & \frac{r_{0i}}{C^2_0}.
\end{eqnarray}
It may be noted that for large $q_i$, the amplitude $t_i$ goes to unity
which is to be expected. The extrapolation of the scattering amplitude
for the off-shell effects can be achieved by means of a monopole form
factor~\cite{lage91}. For a detailed discussion of the off-shell
effects we refer to~\cite{hanh98}.

The factorization of the total amplitude into those of the FSI and primary
production (Eq. (65)), enables one to pursue the diagrammatic approach
for the latter within an effective Lagrangian model and investigate
the role of various meson exchanges and resonances in describing
the reaction. Moreover, in this way the FSI among all the three
outgoing particles can be included. Although the meson-baryon interactions
are weak, they can still be influential through interference. 
 
The parameters $a$ and $r_0$ are very poorly known for the 
$K^+$-nucleon and $K^+$-hyperon systems as the corresponding 
scattering data are scarce at low energies. On the other hand,
for the hyperon-nucleon system several sets of values for these
parameters have been given by the Bonn-J\"ulich~\cite{holl94} and
Nijmegen~\cite{nijm89} groups from their respective $\Lambda - p$
interaction models. There is quite some variation
in the values given in these sets.  For the $K^+$-$p$ and $K^+$-$\Lambda$
systems we have adopted the values given in a recent effective Lagrangian 
model analysis of the available $K\Lambda$ and $Kp$ data by
Feuster and Mosel~\cite{feus98}. In any case, the cross sections are 
insensitive to the FSI effects in these channels. On the other hand,
these effects are very important in $\Lambda-p$ channel, and we 
have performed calculation of the corresponding FSI effects with
all the sets of these parameters given by Bonn-J\"ulich and Nijmegen groups
(given in Table 3) in order to see if the results are sensitive to
various models.  More details will be given in the next section.

\section{Results and discussion}

The theoretical approach presented in the previous section has been used 
to study the available data on the $p$ + $p$ $\rightarrow$ $p$ + $\Lambda$
+ $K^+$ reaction for beam energies ranging from just above the
production threshold to about 10 GeV. In the results shown below, 
we have used PS couplings for both $N^*N\pi$ and $N^*\Lambda K^+$
vertices involving spin-1/2 resonances of even and odd parities.
However, calculations have also been performed with the corresponding
PV couplings. The cross sections calculated with this option for
the resonance-hyperon-kaon vertex deviate very little from those 
obtained with the corresponding PS couplings. However, the PV coupling
for the $N^*N\pi$ vertex leads to noticeably different results as is
discussed below.

\subsection{Cross section data for beam energy above 2 GeV}

In Fig. 2 we show the comparison
of our calculations with the experimental data for the total 
cross section for this reaction as a function of beam momentum for 
incident energies above 2 GeV. In this
figure we have investigated the role of various meson exchange processes
in describing the total cross section. The dotted, dashed, long-dashed and 
dashed-dotted curves represent the contributions of $\pi$, $\rho$, $\omega$
and $\sigma$ meson exchanges respectively. The contribution of the 
heavy axial meson exchange is not shown in this figure as it is  
negligibly small. The coherent sum of all the meson-exchange processes
is shown by the solid line. The experimental points are taken
from~\cite{land88}. We notice that the measured cross sections
are reproduced reasonably well by our calculations (solid line) for
all the beam
energies except for two lowest points. The FSI 
effects which are not included in these calculations, reduce the
discrepancy between the experimental data and calculations at 
these beam momenta. This point is further discussed in the next subsection. 
 
\vbox{  
\begin{table}[here]
\begin{center}
\caption { Scattering length ($a$) and effective range ($r_0$) 
parameters for the $\Lambda N$ scattering derived from models
$A$, ${\tilde A}$, $B$ and ${\tilde B}$ of the J\"ulich-Bonn group
\protect\cite{holl94}
and $D$, $F$ and $NSC$ of the Nijmegen group~\protect\cite{nijm89} }
\vspace{0.5cm}
\begin{tabular}{|c|c|c|c|c|} \hline
model       &$a$(singlet) &$r_0$(singlet) & $a$(triplet) & $r_0$(triplet) \\
            & (fm)       & (fm)          &(fm)          &(fm)   \\ \hline
$A$           & 1.56        & 1.43          & 1.59         & 3.16   \\
${\tilde A}$ & 2.04        & 0.64          & 1.33         & 3.91   \\
$B$           & 0.56        & 7.77          & 1.91         & 2.43    \\
${\tilde B}$ & 0.40        & 12.28         & 2.12         & 2.57    \\
$D$           & 1.90        & 3.72          & 1.96         & 3.24    \\ 
$F$           & 2.29        & 3.17          & 1.88         & 3.36     \\
$NSC$         & 2.78        & 2.88          & 1.41         & 3.11    \\
\hline 
\end{tabular}
\end{center}
\end{table}
}

We note that the pion exchange graphs dominate the production process
for all the energies. In comparison to this, the contributions of 
$\rho$ and $\omega$ meson exchanges are almost insignificant.
The $\rho$-meson exchange, which is a convenient way of taking
into account the $P$ wave part of the correlated two-pion exchange (CTPE) 
process, is rather week even  in the low energy
$NN$ scattering~\cite{bonn87}. 
With increasing projectile energy its contribution decreases further.
On the other hand, the $\sigma$ meson exchange, which models the CTPE
in the $\pi\pi$ S-wave and provides about 2/3 of this exchange in
the low energy $NN$ interaction, plays a relatively more important role.
This observation has also been made in the case of
$NN\rightarrow NN\pi$ reaction~\cite{dmit86,risk93,horo94,enge96,shya98}.
Thus, the $\sigma$ meson exchange provides an efficient
means of mediating the large
momentum mismatch involved in the meson production reactions in $NN$
collisions, particularly at lower beam momenta.
 
The relative importance of the contributions of each intermediate
resonance to the $pp\rightarrow p\Lambda K^+$ reaction is studied in
Fig. 3, where the contributions of 
$N^*$(1650), $N^*$(1710) and $N^*$(1720) resonances to the energy
dependence of the total cross section are shown by dotted, dashed
and dashed-dotted lines respectively. Their coherent sum is depicted 
by the solid line. It is clear that the contributions
from the $N^*$(1710) and $N^*$(1650) resonances dominate the total cross
section at beam momenta above and below 3 GeV/c respectively. Moreover,
the interference terms of the amplitudes corresponding to various resonances
are quite important. This result is in sharp contrast to the 
resonance model calculations of Refs.~\cite{tsus94,tsus97,sibi98}, where
these terms were ignored. It must again be emphasized that we have
absolutely no freedom in choosing the relative signs of
the interference terms.
 
\begin{figure}[here]
\begin{center}
\epsfxsize=8.2cm
\epsfbox{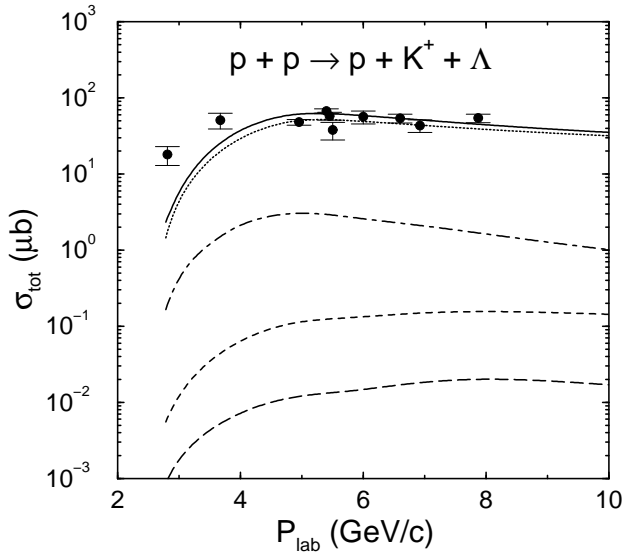}
\end{center}
\caption {
The total cross section for the
$p+p \rightarrow p+K^{+}+\Lambda$
reaction as a function of the beam momentum. The dotted, dashed,
long-dashed and dashed-dotted curves represent the contributions of
$\pi$, $\rho$, $\omega$ and $\sigma$ meson exchanges respectively. Their
coherent sum is shown by the solid line. The experimental data are
from~\protect\cite{land88}.} 
\label{fig:figb}
\end{figure}
 
Looking in Table 2, one might  
naively expect the dominance of the $N^*$(1710) resonance everywhere as 
the coupling constants for the $N^*\Lambda K^+$ and $N^*N\pi$ vertices
for the  $N^*$(1710) resonance are  about an order of magnitude larger
than those for $N^*$(1720) and $N^*$(1650) resonances. In fact, the 
relative importance of various resonances is determined by the dynamics
of the reaction mechanism. As the beam energy rises above the $K^+$ 
production threshold, the excitation of the resonance lowest in energy
is more probable in the beginning. However, with increasing beam energy
the excitation of the higher energy resonances starts playing
more and more important role.

As mentioned earlier, the use of the PV coupling for the $N^*\Lambda K$
vertices (involving spin-1/2 even and odd parity resonances) makes
insignificant changes in the cross sections. However, there
is a clear preference for the PS coupling at the $N^*N\pi$ vertices.
This is shown in Fig. 4, where the
ratio of the total cross section obtained by using the PV ($\sigma_{PV}$)
and PS ($\sigma_{PS}$) couplings for these vertices, is shown as a 
function of the beam momentum. It is seen that $\sigma_{PV}$ is 
larger than $\sigma_{PS}$ at higher beam momenta while at lower ones
the reverse is true. Clearly PS coupling for the $N^*N\pi$ vertex 
provides a better description of the beam energy dependence of the
total cross section for the $pp \rightarrow pK^+\Lambda$ reaction. 

\begin{figure}[here]
\begin{center}
\epsfxsize=8.2cm
\epsfbox{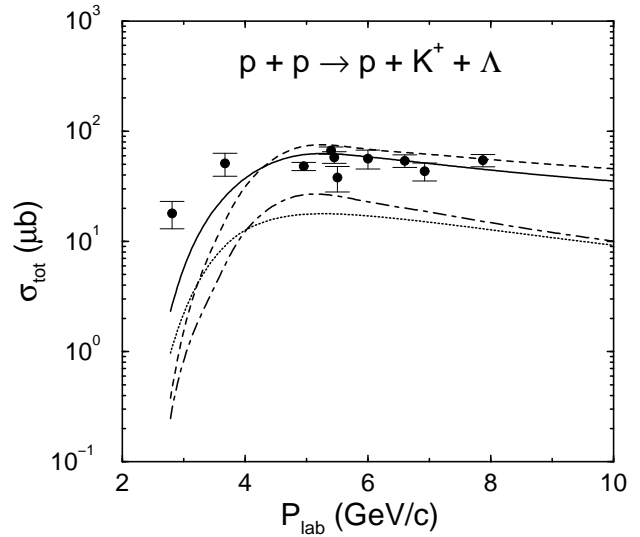}
\end{center}
\caption {
Contributions of N$^{*}$(1650) (dotted line), N$^{*}$(1710)
(dashed line) and N$^{*}$(1720) (dashed-dotted line) baryonic resonances
to the total cross section for the 
$p+p \rightarrow p+K^{+}+\Lambda$ reaction as a function of beam momentum. 
Their coherent sum is shown by the solid line.} 
\label{fig:figc}
\end{figure}

\subsection{Cross section data for beam energies below 2 GeV}

In Fig. 5 we compare the results of our calculations (with FSI effects
included) with the recent data~\cite{juel98} for the $pp \rightarrow pK^+
\Lambda$ reactions at beam energies very close to the kaon production
threshold. In this figure the total cross section is shown as a 
function of the excess energy ($\epsilon$) $=$ $\sqrt{s}$ - $m_N$ - $m_{K^+}$
- $m_{\Lambda}$, where $\sqrt{s}$ is the invariant mass. The FSI
effects were included by following the 
procedure outlined in section 2.G.  We have chosen~\cite{feus98}
$a$ = $0.065+i0.040$, $r_0$ = $-15.930-i8.252$ and $a$ = $-0.214$,
$r_0$ = $-0.331$ for the $K^+\Lambda$ and $K^+p$ systems respectively in
all the calculations shown below. For the $\Lambda-p$ system all the
seven sets of the parameters as shown in Table 3 were used.
 
In Fig. 5a, the results obtained with the parameters sets of models 
$A$ (dotted line), ${\tilde {A}}$ (solid line), $B$ (dashed line) and
${\tilde {B}}$ (long-dashed-dotted line) of the
Bonn-J\"ulich group~\cite{holl94} are shown. It can be noted that
all the four models provide similar results for the total cross
sections at $\epsilon$ $\sim$ 150 MeV. However, at lower values
of $\epsilon$, cross section calculated with models $A$, and ${\tilde {A}}$
are larger than those of model $B$ and ${\tilde {B}}$ by a factor
of about 2-3. Moreover, there is a difference of a factor of more than
2 between the results obtained with model $A$ and ${\tilde {A}}$ itself
with the latter providing a better overall
agreement with the data. We also show in this figure the results 
obtained without including the FSI effects (dashed-dotted line). It
can be noted that the FSI effects are quite important in order to describe
the experimental data.

\begin{figure}[here]
\begin{center}
\epsfxsize=8.2cm
\epsfbox{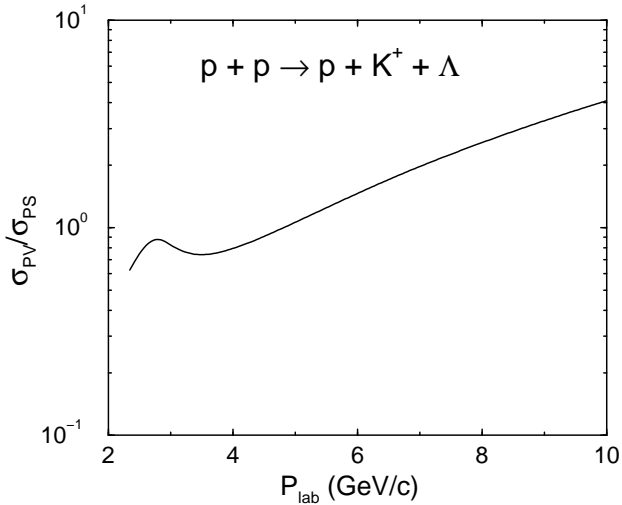}
\end{center}
\caption{ Ratio of the total cross section calculated with
pseudovector and pseudoscalar couplings for the $N^*N\pi$ vertex 
corresponding to  spin-1/2 (even and odd parity) resonance for the
same reaction as in Fig.2, as a function of beam momentum.}
\label{fig:figd}
\end{figure}
 
The results obtained with models $D$, $E$ and
$NSC$ of the Nijmegen group~\cite{nijm89} are shown in Fig. 5b. These
three models produce almost identical results for all the values
of $\epsilon$. However, while the data at the higher excess energies 
are reproduced by all the three model quite well,  they under-predict
the cross sections at lower $\epsilon$ by a factor of about 3.
Therefore, while all the models of $\Lambda-p$ interaction
(considered in this work) provide an equally good description
of the total cross section data at higher values of the excess
energy, a difference of factors of 2-3 occurs between
their predictions at lower values of $\epsilon$. Thus, the near threshold 
$\Lambda K^+$ production data in proton-proton collisions are sensitive 
to the S-wave $\Lambda$-nucleon interaction and may be used
to distinguish between various models proposed in the literature to
describe this interaction. We note that model ${\tilde A}$ of the
Bonn-J\"ulich group provides the best overall description of
the data, which has been used to account for the $\Lambda - p$ FSI
effects in all the calculations discussed subsequently.

The individual contributions of various nucleon resonances to the 
total cross section of the $pp \rightarrow p\Lambda K^+$ reaction
near the production threshold is shown in Fig. 6, as a function of
the excess energy. In contrast to the situation at higher beam energies
($p_{lab}$ $\geq$ 3 GeV/c), the cross section is dominated by the
$N^*$(1650) resonance excitation. This is in line with the observations 
made in Refs.~\cite{fald97,sibi98}.
It may, however, be noted that in Ref.~\cite{sibi98}, 
the FSI effects have not been included and no comparison with the data
at near threshold energies is shown.

\begin{figure}[here]
\begin{center}
\epsfxsize=8.2cm
\epsfbox{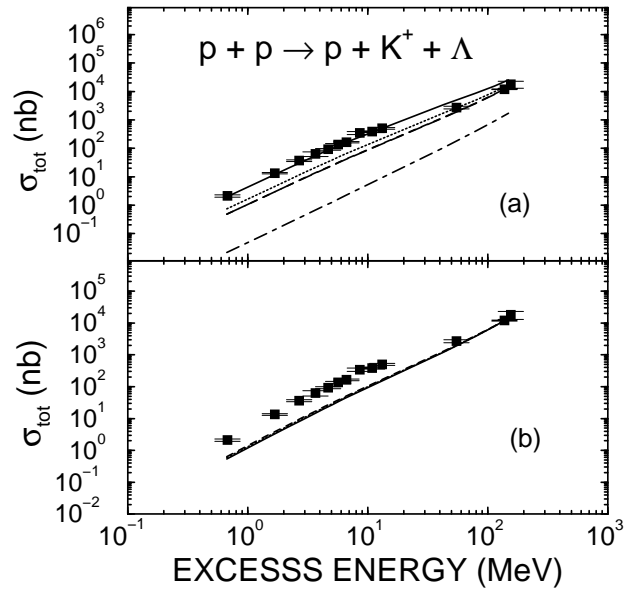}
\end{center}
\caption {
The total cross section for the
$p+p \rightarrow p+K^{+}+\Lambda$ reaction very close to the $K^{+}$ 
production threshold as a function of the excess energy (defined in the
text). The FSI effects are included by using the scattering length ($a$) and
effective range ($r_0$) parameters for the $K^+-\Lambda$ and $K^+-p$
systems taken from the Ref.~\protect\cite{feus98}
and those for the $\Lambda -p$ 
system from the sets given by  J\"ulich-Bonn~\protect\cite{holl94}  
and Nijmegen~\protect\cite{nijm89}
groups. In the upper part (a) results obtained
with the $\Lambda-p$ parameters of models A (dotted line),
${\tilde A}$ (solid line), B (dashed line) and ${\tilde B}$
(long-dashed line) of the former group are shown, while in the
lower part (b) those of models D, F and NSC of
the latter group are depicted. Results of the three models of the
Nijmegen group are indistinguishable from each other. In the upper part
(a) results with no FSI effects are shown by dashed-dotted line. The
experimental data are taken from~\protect\cite{land88,juel98}.} 
\label{fig:fige}
\end{figure}

Since $N^*$(1650) is the lowest energy baryonic
resonance having an appreciable branching ratio for the decay to the
$\Lambda K^+$ channel, its dominance in this reaction at beam energies 
near the kaon production threshold is to be expected. The contributions
of other two resonances ($N^*$(1710) and $N^*$(1720)) are several orders
of magnitude smaller, therefore, the resonance-resonance interference
terms are also very small. Thus, near threshold energies, this reaction
proceeds preferentially via excitation of the $N^*$(1650) resonance.
It may be noted that in Ref.~\cite{sibi98}, 
the FSI effects have not been included in
these calculations.
 
In Fig. 7, we show the contributions of various meson exchanges to this
reaction at near threshold beam energies. Various curves have the
same meaning as in Fig. 2. The one pion exchange graphs dominate the
reaction in this energy regime as well. On the other hand, 
the individual contributions of the $\rho$ and $\omega$ meson
exchange processes are negligible. However, those of the 
$\sigma$ meson exchange are substantial in this
energy regime. Thus, like near threshold pion production in proton-proton
collisions, the heavy scalar meson exchange plays an important
role in this case too.  It should however, be noted that the interference
terms of various meson exchange processes are not negligible; contributions
of various exchange processes simply do not add up to the total cross
section obtained by the coherent addition of various amplitudes. 
 
\begin{figure}[here]
\begin{center}
\epsfxsize=8.2cm
\epsfbox{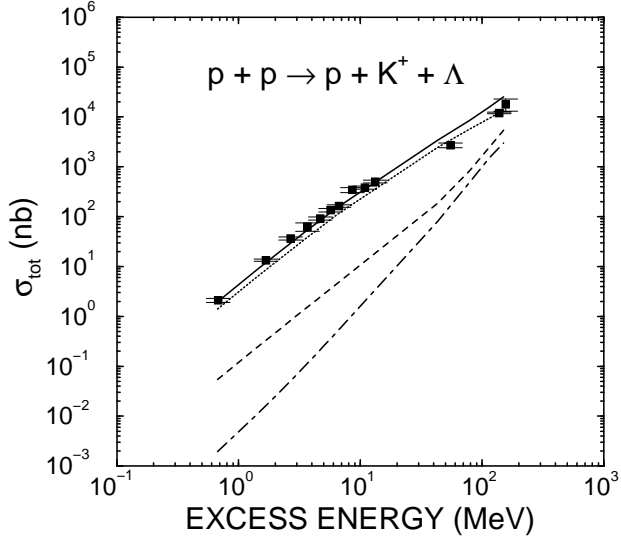}
\end{center}
\caption {
Contributions of N$^{*}$(1650) (dotted line), N$^{*}$(1710)
(dashed line) and N$^{*}$(1720) (dashed-dotted line) baryonic resonances
to the total cross section for the same reaction as in Fig. 5, 
as a function of the excess energy. Their coherent sum is shown by the solid
line. The FSI effects are included with $a$ and $r_0$ parameters of the 
$K^+-p$ and $K^+-\Lambda$ systems being the same as those in Fig. 5  
and those for the $\Lambda - p$ system being taken from model ${\tilde A}$
of the J\"ulich-Bonn group~\protect\cite{holl94}.} 
\label{fig:figf}
\end{figure}

\section{Summary and conclusions}

We investigated the associated $K^+\Lambda$ production in the proton-proton
collisions at energies ranging from near threshold to about 10 GeV. This
reaction is of interest as it provides the prospect of testing QCD in the
non-perturbative domain at energies larger than the pion mass. In this 
paper our goal has been to investigate this reaction within an effective
Lagrangian model which is proven to be very successful in describing the
pion production in $NN$ collisions. Most of the parameters of this 
model are fixed by fitting to the elastic $NN$ T-matrix; this restricts 
the freedom of varying the parameters of the model to provide a fit to 
the data.

The reaction proceeds via the excitation of the
$N^*$(1650), $N^*$(1710)
and $N^*$(1720) intermediate nucleon resonant states. The coupling constants
at vertices involving resonances have been determined from the 
experimental branching ratios of their decay into various relevant channels.
Unlike the $NN\pi$ vertex, there is no compelling reason to choose the 
pseudovector (PV) form for the $N^*\Lambda K^+$ and $N^*N\pi$ couplings
(involving spin-1/2 resonances of even and odd parities) and we investigated
both the PV and pseudoscalar (PS) couplings at these vertices. 
To describe the data at the near threshold beam energies, we have included
the FSI effects among the outgoing particles by following Watson-Migdal 
theory which has been used before successfully to describe the $NN\eta$
and $NN\pi$ reactions in the similar energy regimes. 

\begin{figure}[here]
\begin{center}
\epsfxsize=8.2cm
\epsfbox{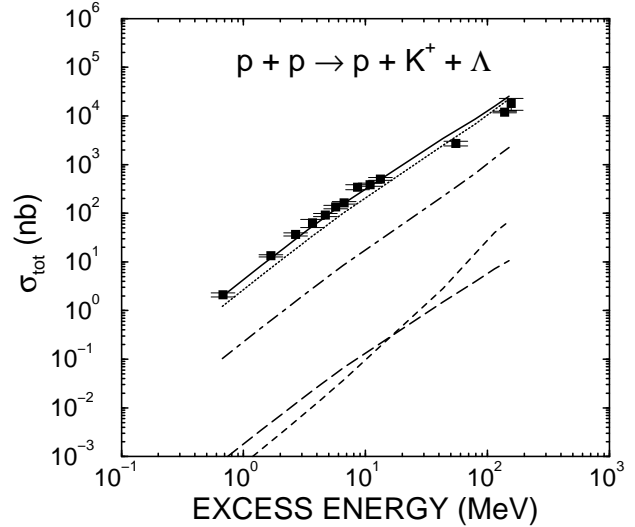}
\end{center}
\vskip .1in
\caption {
Contributions of $\pi$ (dotted line), $\rho$
(dashed line), $\omega$ (long-dashed line) and $\sigma$ (dashed-dotted line)
meson exchange processes to the total cross section for the 
same reaction as shown in Fig. 6, as a function of the excess energy.
Their coherent sum is shown by the solid line. The FSI effects
are included in the same way as in Fig. 6. } 
\label{fig:figg}
\end{figure}

With the same set of parameters, the model is able to provide a good
description of the data at higher as well as near threshold beam energies.
The one-pion-exchange processes make the dominant contribution to the
cross section in the entire energy regime. The individual contributions
of the $\rho$ and $\omega$ meson exchange diagrams is very small every where.
Although, the interference terms of their amplitudes with those of 
other meson exchanges may  still be noticeable. 
On the other hand, the $\sigma$ exchange makes a
relatively larger contribution at lower beam energies, confirming the
earlier observation that the heavy scalar meson provides a means of
mediating the large momentum transfer in near threshold $NN$-meson
production processes. 

While at beam momenta larger than 3 GeV/c, the reaction proceeds predominantly
via excitation of the $N^*$(1710) resonance, the process gets maximum 
contribution from the $N^*$(1650) resonance at lower beam energies. A 
very striking feature of our results is that the interference among various
resonance contributions is very significant. Therefore, in the calculations 
of this reaction these terms should not be ignored.  
The near threshold data clearly favors the excitation of the
$N^*$(1650) resonance. Therefore, this reaction, in this
energy regime, provides an ideal means of investigating
the properties of this baryonic resonance.

Unlike the $NN\pi$ vertex where there is a clear preference for the
PV coupling, as seen in the $NN\pi$ data, the present reaction does
not distinguish between PS and PV couplings at the $N^*\Lambda K^+$
vertex involving spin-1/2 even or odd parity resonance. However,
the PS coupling at the $N^*N\pi$ vertex is clearly favored by the data.

The near threshold data may be selective about the model describing the
low energy $\Lambda$-nucleon scattering. Calculations of the  
FSI effects performed with the scattering length and effective range 
parameters of the J\"ulich-Bonn group produce different results as 
compared to those performed with the corresponding parameters of the
Nijmegen group. The parameters of model ${\tilde A}$ of
Ref.~\cite{holl94} provides the best agreement with the data.

An obvious extension of the present work is to calculate the cross sections 
for the $pp \rightarrow p\Sigma K^+$ reaction, for which the measurements
have recently been performed at COSY~\cite{sewe98}. This will also lead to 
the inclusive $K^+$ cross sections in the elementary
nucleon-nucleon collisions which are the necessary input to the transport
model calculations of the strangeness production in the heavy ion collisions.
This work is currently underway by extending the model to include the
excitation of delta isobars $\Delta$(1910) and $\Delta$(1920) which are
four star and three star resonances respectively. Since, 
the on-shell $N^*$(1650) $\rightarrow \Sigma K$ decay is not allowed,
techniques similar to those described in section (D.ii) will
have to be used to calculated the coupling constant for this 
vertex. It would also be interesting to calculate the invariant
mass spectrum of the hyperon $K^+$ pair which is expected to provide 
further information about the various resonance contributions to 
this reaction. This will be reported in a future publication.
Extensions of the present theory to incorporate the unitarity, perhaps 
on the lines of the $K$-matrix approximation~\cite{feus98}
is also desirable. 

\acknowledgements

The author is thankful to Ulrich Mosel for his very kind 
hospitality in the University of Giessen during several visits
and for numerous useful discussions which were very helpful  
in completing this work.  He also wishes to thank Wolfgang N\"orenberg
and J\"orn Knoll for their warm hospitality in the theory group
of the GSI where a part of this work was done. Useful discussions
with W. Cassing, B. Friman, G. Penner, W. Peters, M. Post and
A. Sibirtsev are gratefully acknowledged. The author would like
to acknowledge the financial support from the Abdus Salam International
Center for Theoretical Physics, Trieste.

\end{document}